\documentclass[preprint,aps,showpacs]{revtex4}
\usepackage{amsfonts}
\usepackage{amsmath}
\usepackage{amssymb}
\usepackage{graphicx}

\newcommand{\be}{\begin{equation}}
\newcommand{\ee}{\end{equation}}
\newcommand{\bes}{\begin{subequations}}
\newcommand{\ees}{\end{subequations}}
\newcommand{\ben}{\begin{eqnarray}}
\newcommand{\een}{\end{eqnarray}}

\begin{document}

\title{Extended solutions via the trial-orbit method for two-field models}
\author{A.R. Gomes$^a$ and D. Bazeia$^b$ }
\affiliation{{\small {$^a$Departamento de F\' isica, Centro Federal de Educa\c
c\~ao Tecnol\'ogica do Maranh\~ao, Brazil \footnote{e-mail: argomes@pq.cnpq.br, tel: xx-55-98-32351384}}\\
$^b$Departamento de F\'\i sica, Universidade Federal da Para\'\i ba, Brazil}}

\begin{abstract}
In this work we investigate the presence of defect structures in models
described by two real scalar fields. The coupling between the two fields is
inspired on the equations for a multimode laser, and the minimum energy
trivial configurations are shown to be structurely dependent on the
parameters of the models. The trial orbit method is then used and several
non-trivial analytical solutions corresponding to topological solitons are
obtained.
\end{abstract}


\maketitle



The study of topological defects is a well established field, particularly
for models described by scalar fields \cite{raj1,msu}. The simplest
topological defect - the kink - arises in theories of scalar fields in
two-dimensional space-time\cite{rub}. For usual models with spontaneous
breaking of global symmetry, such defects interpolate between two minima of
the potential. Important examples in condensed matter physics are the well-known
domain walls, which separate regions of different magnetization. These
defects are essentially classical objects with localized and stable
distribution of density energy. In the case of two coupled real scalar fields, the
equations of motion are very hard to solve due to non-trivial
nonlinearities. However, there are interesting situations where real
progress have been done -- see, e.g. Refs.\cite{t1,t2,t3,t4,t5,t6}.

In 1979, Rajaraman proposed a method to solve the pair of equations of motion
which usually appear in models described by two real scalar fields \cite{t4}%
: it is named the trial orbit method, which relies on the search (in a trial
way approach) for an appropriate orbit the two fields have to obey in the
two-dimensional configuration space. Eventually, when one tries the right
orbit, we can be able to solve the problem analytically. However, since the
equations of motion are second order differential equations, the task of
finding exact solutions is very hard and the trial orbit method is not 
much efficient.

Some years before - in 1976 - an interesting work \cite{bog} identified an
important class of models, showing how to reduce the equations of motion to a
system of first order differential equations. In 1995, this subject was studied by 
one of us in ref. \cite{bsr}, that is, the Rajaraman's trial orbit method 
\cite{t4} was applied for the first order equations obtained within the Bogomol'nyi
procedure \cite{bog}. The use of the trial orbit method for first order
differential equations was shown to be very efficient and this new procedure
allowed us to make interesting progress, as it is shown in \cite{bb}
and in references therein. 
More recently the use of the trial orbit
method for models whose equations of motion can be reduced to first
order differential equations was systematized in \cite{bflr}. Other investigations on similar issues have also been done in \cite{refb}-\cite{d}, which use distinct procedures and motivations to study two-field and other related models.

In the case of a model with two fields, the kink-like solutions are orbits
in the field space. In this work, we will further explore the trial orbit
method to investigate models described by first order equations. Here,
although, we construct a class of models inspired in a semiclassical theory
of multimode laser and use the trial orbit method to find exact solutions
that minimize the energy of the field configurations. The results  show
that, under certain conditions on the parameters of the system, 
several possible solutions connecting distinct minima of the models exist.

We consider a class of models in (1,1) Minkowski space-time dimensions described by
the relativistic Lagrange density 
\begin{eqnarray}
{\mathcal{L}}=\frac{1}{2}\partial_{\mu}\phi_1\partial^{\mu}\phi_1+\frac{1}{2}%
\partial_{\mu}\phi_2\partial^{\mu}\phi_2-V(\phi_1,\phi_2)
\end{eqnarray}
where $\phi_1$ and $\phi_2$ are the two real scalar fields, and we use the
metric such that $x^0=x_0=t$ stands for the time, while $x^1=-x_1=x$
represents the spatial coordinate. The notation is usual for relativistic
theories, with upper (lower) $\mu$ standing for contravariant (covariant)
coordinates. The metric tensor is a diagonal $2\times2$ matrix, compactly 
written as $g_{\mu\nu}=(1,-1)$. The Euler-Lagrange equation, 
\ben
\partial_\mu\frac{\partial\mathcal{L}}{\partial(\partial_\mu\phi)}-\frac{\partial\mathcal{L}}{\partial\phi}=0
\een
leads to the following equations of motion:
\begin{subequations}
\begin{eqnarray}
\frac{\partial^2\phi_1}{\partial t^2}-\frac{\partial^2\phi_1}{\partial x^2}+%
\frac{\partial V}{\partial\phi_1}&=&0 \\
\frac{\partial^2\phi_2}{\partial t^2}-\frac{\partial^2\phi_2}{\partial x^2}+%
\frac{\partial V}{\partial\phi_2}&=&0
\end{eqnarray}%
We are interested in kink-like solutions, which are described by static
fields - $\phi_1=\phi_1(x)$, $\phi_2=\phi_2(x)$ - so that 
\end{subequations}
\begin{eqnarray}
\label{second_eom}
\frac{d^2\phi_1}{dx^2}=\frac{\partial V}{\partial\phi_1};
\;\;\;\;\;\;\;\;\;\; \frac{d^2\phi_2}{dx^2}=\frac{\partial V}{\partial\phi_2}
\end{eqnarray}
In general, these equations are very hard to solve, but this task may be simplified 
if it is possible to replace these second order equations by
first order differential equations. In order to get first order
equations, we suppose that the potential is given in terms of another
function, $W=W(\phi_1,\phi_2),$ as bellow: 
\begin{eqnarray}
V(\phi_1,\phi_2)=\frac12\left(\frac{\partial W}{\partial\phi_1}%
\right)^2+\frac12\left(\frac{\partial W}{\partial\phi_2}\right)^2
\end{eqnarray}
In this case, the Bogomol'nyi method allows to argue that the solutions of the first order equations 
\begin{eqnarray}
\frac{d\phi_1}{dx}=\frac{\partial W}{\partial\phi_1}; \;\;\;\;\;\;\;\;\;\; 
\frac{d\phi_2}{dx}=\frac{\partial W}{\partial\phi_2}
\end{eqnarray}
are also solutions of Eqs. (\ref{second_eom}), as it can be easily verified.

The potential of the above model has zeroes at the singular points of $%
W(\phi_1,\phi_2)$, and this set of singular points forms the vacua manifold
of the field theory under investigation. Usually, distinct pairs of
minima define distinct topological sectors of the model, and the solutions
of the first order equations are defect structures with an energy cost 
given by $E=|\Delta W|,$ where 
\begin{eqnarray}
\Delta W=W(\phi_1(+\infty),\phi_2(+\infty)) -
W(\phi_1(-\infty),\phi_2(-\infty))
\end{eqnarray}
with the points $(\phi_1(+\infty),\phi_2(+\infty))$ and $(\phi_1(-\infty),%
\phi_2(-\infty))$ identifying minima in the vacua manifold. Since the energy
density of the static fields is given by 
\begin{eqnarray}
\epsilon(x)=\frac12\left(\frac{d\phi_1}{dx}\right)^2+\frac12\left(\frac{%
d\phi_2}{dx}\right)^2+ \frac12\left(\frac{\partial W}{\partial\phi_1}%
\right)^2+\frac12\left(\frac{\partial W}{\partial\phi_2}\right)^2
\end{eqnarray}
the energy is always positive, and the solutions which obey
the first order equations are the minimum energy configurations in each
topological sector of the model.


To be specific, let us now consider the superpotential 
\begin{equation}
W=\frac{1}{2}\mu _{1}\phi _{1}^{2}+\frac{1}{2}\mu _{2}\phi _{2}^{2}-\frac{1}{%
4}\lambda _{11}\phi _{1}^{4}-\frac{1}{4}\lambda _{22}\phi _{2}^{4}-\frac{1}{2%
}\lambda _{12}\phi _{1}^{2}\phi _{2}^{2}
\label{Wfichi}
\end{equation}%
This choice represents a class of models described by the two sets of
parameters: $\{\mu _{1},\mu _{2}\},$  and $%
\{\lambda _{11},\lambda _{22},\lambda _{12}\},$ the first being mass parameters while 
the second specifying interactions
between the two fields. This potential implies the following first order differential equations: 
\begin{subequations}
\label{genmoveq}
\begin{eqnarray}
\frac{d\phi _{1}}{dx} &=&(\mu _{1}-\lambda _{11}\phi _{1}^{2}-\lambda
_{12}\phi _{2}^{2})\phi _{1}  \label{genmoveq1} \\
\frac{d\phi _{2}}{dx} &=&(\mu _{2}-\lambda _{21}\phi _{1}^{2}-\lambda
_{22}\phi _{2}^{2})\phi _{2}  \label{genmoveq2}
\end{eqnarray}%
where we have set $\lambda _{21}=\lambda _{12}$.

The present model represents in reality a family of models which is refereed to
it some generality. Moreover, there is another specific motivation to adopt it:
the system of Eqs. (\ref{genmoveq}) is  connected with the semiclassical
theory of the laser and can simulate the competition between two adjacent
modes in a cavity above the threshold (\cite{nuss},
pp.126-131). It is said that the laser is at threshold when the pumping rate
from the lower state to the upper excited state is just sufficient to overcome the 
cavity loss. 
In this way, 
for the particular case of a two-mode laser, within the approximation that the induced transition rate is well below the saturation rate, we have (note the resemblance with Eqs. (\ref{genmoveq}a)-(\ref{genmoveq}b)):
\ben
\label{eqEn}
\dot{E_n}=\mu_nE_n-\lambda_{nn}E_n^3-\sum_{m\neq m}\lambda_{nm}E_nE_m^2, n=1,2
\een
Here $E_n$ is the time-dependent slow-varying amplitude associated with the mode n, after expanding the electric field in the cavity in terms of a complete set of axial modes.   
With this 
motivation, the parameters $\mu_1$ and $\mu_2$ represent the
overall gain, with the condition $\mu_i\geq 0$ being necessary to establish the laser 
oscillation in the mode i ($i=1,2$). Furthermore, $\lambda_{11}$ and $\lambda_{22}$ are
saturation parameters, and one must have $\lambda_{ii}>0$ for positive
population inversion of the mode $i$. The parameter $\lambda_{12}$ stands
for the nonlinear saturation effect on the coupling between the two modes.
We also have for the two-mode laser $\lambda_{12}\sim\lambda_{21}>0$. The study of competition among modes considers the analysis of the stability of the stationary solutions in a phase space diagram of $E_1^2$ versus $E_2^2$, where numerical solutions for arbitrary initial conditions reveal the stable and unstable points. It is found that stability of solutions is strongly dependent on the parameters, where one can have laser oscillation in just one of the modes or a simultaneous oscillation is both modes.  

Our work considers a similar problem. However, instead of investigate 
$\phi _{1}^{2}$ and $\phi _{2}^{2}$ in a phase space diagram, we follow another route and make an analysis in
connection with the field description, searching for analytical description
of the fields $\phi_1$ and $\phi_2$. To make the work as general as possible, let us start
considering the vacua manifold, e.g. searching for all the possible minimum
energy points of the potential, the critical points of $W.$ Initially we can
count five points $(\phi _{1},\phi _{2})$ of minima: $(0,0),(\pm \phi
_{1}^{\ast },0)$,$(0,\pm \phi _{2}^{\ast })$ with $\phi _{1}^{\ast }=\sqrt{%
\mu _{1}/\lambda _{11}}$ and $\phi _{2}^{\ast }=\sqrt{\mu _{2}/\lambda _{22}}
$ -- see Fig.~1. The case where both $\phi _{1}$ and $\phi _{2}$ are non
vanishing can lead to 4 more points of minima, a continuum of points or no
more points, depending on the relation between the parameters. We use the
first order equations \eqref{genmoveq} to get 
\end{subequations}
\begin{subequations}
\label{elipse}
\begin{eqnarray}
\mu _{1}-\lambda _{11}\phi _{1}^{2}-\lambda _{12}\phi _{2}^{2} &=&0
\label{elipse1} \\
\mu _{2}-\lambda _{21}\phi _{1}^{2}-\lambda _{22}\phi _{2}^{2} &=&0
\label{elipse2}
\end{eqnarray}%
We then define the matrices 
\end{subequations}
\begin{equation}
\Lambda \equiv \left( 
\begin{array}{cc}
\lambda _{11} & \lambda _{12} \\ 
\lambda _{21} & \lambda _{22} \\ 
& 
\end{array}%
\right) \;\;\;\;\;\Lambda ^{(\phi _{1})}\equiv \left( 
\begin{array}{cc}
\mu _{1} & \lambda _{12} \\ 
\mu _{2} & \lambda _{22} \\ 
& 
\end{array}%
\right) \;\;\;\;\;\Lambda ^{(\phi _{2})}\equiv \left( 
\begin{array}{cc}
\lambda _{11} & \mu _{1} \\ 
\lambda _{21} & \mu _{2} \\ 
& 
\end{array}%
\right) 
\end{equation}%
We can analyze better the structure of the solutions expressing the former
equations in a matricial form $\Lambda \overrightarrow{\Phi ^{2}}=\vec{\mu}$.

$\bullet$ For $\det(\Lambda)\neq 0$ we have a formal solution $%
\overrightarrow{\Phi^2}={\Lambda^{-1}}\vec{\mu}$ and the four minima $%
(\pm\bar\phi_1,\pm\bar\phi_2)$, with 
\begin{equation}
\bar\phi_1=\sqrt\frac{\mu_2\lambda_{12}-\mu_1\lambda_{22}} {%
\lambda_{12}^2-\lambda_{11}\lambda_{22}}  \label{barphi}
\end{equation}
\begin{equation}
\bar\phi_2=\sqrt\frac{\mu_1\lambda_{21}-\mu_2\lambda_{11}} {%
\lambda_{21}^2-\lambda_{11}\lambda_{22}}  \label{barchi}
\end{equation}
See Fig.1(a).

$\bullet$ For $\det(\Lambda)=\det(\Lambda^{(\phi_1)})=\det(\Lambda^{(%
\phi_2)})=0$ we have $\lambda_{12}=\pm\sqrt{\lambda_{11}\lambda_{22}}$. This
means coalescence between the ellipses represented by Eqs. (\ref{elipse1})
and (\ref{elipse2}) and we have an infinity of solutions. See Fig.~1(c).

$\bullet$ For $\det(\Lambda)=0;
\det(\Lambda^{(\phi_1)}),\det(\Lambda^{(\phi_2)})\neq0$ there are no 
solutions satisfying both Eqs. (\ref{elipse1}) and (\ref{elipse2})  and we
have a situation of non-touching ellipses. See Fig.~1(b).

\begin{figure}[ht]
\begin{center}
\includegraphics[height=20cm]{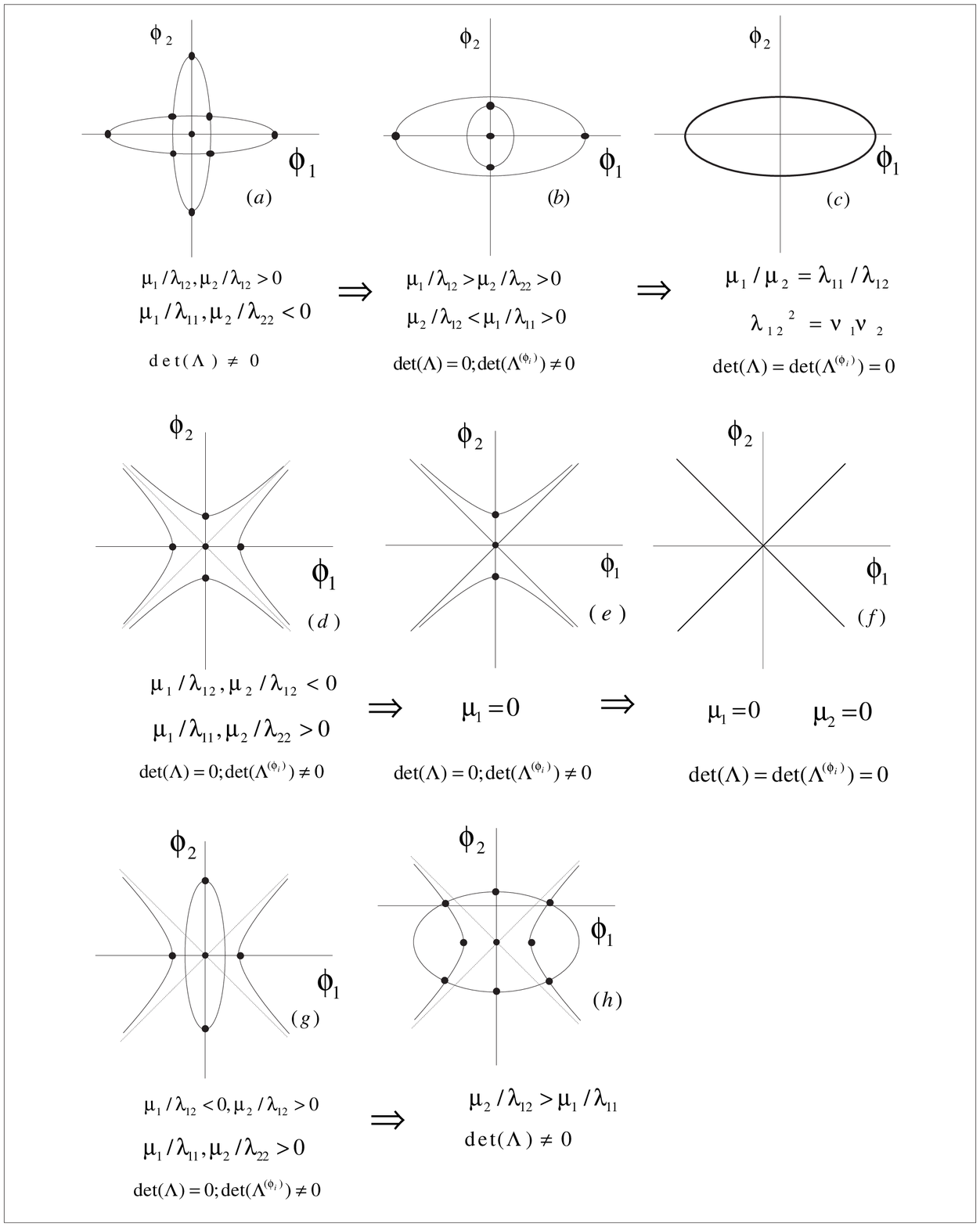}
\end{center}
\caption{Diagrams showing all the possible minimum energy points.}
\label{alldiagrams}
\end{figure}

There are other possibilities, which are also shown in Fig. \ref{alldiagrams}%
. In the diagrams depicted in Fig.1, we show how the minimum energy points
change with the signal of the fractions $\mu_1/\lambda_{12}$, $%
\mu_2/\lambda_{12}$, $\mu_1/\lambda_{11}$ and $\mu_2/\lambda_{22}$. In the
following we analyze solutions connecting pairs of minima related to the
configurations shown in this figure.

We first deal with the case involving the two crossing lines of minima, as
depicted in Fig.~1(f). We use equations \eqref{elipse} to get 
\begin{equation}
{\bar\phi_1}^{2} = -\frac {\lambda_{12}} {\lambda_{11}}{{\bar\phi_2}^2},
\,\,\,\, {\bar\phi_1}^{2} = -\frac {\lambda_{22}}{\lambda_{21}} {{\bar\phi_2}%
^2}  \label{fi1}
\end{equation}
These expressions lead to $\lambda_{12} = \pm\sqrt{\lambda_{11}\lambda_{22}}$%
. Now for $\bar\phi_1$, $\bar\phi_2 \neq0$ we have ${\lambda_{12}}/{%
\lambda_{11}}<0$ and ${\lambda_{22}}/{\lambda_{21}}<0$. Then we have the
following choices: (a) if $\lambda_{11}<0 \implies \lambda_{12}>0$ and $%
\lambda_{22}<0$, or if $\lambda_{22}<0 \implies \lambda_{21}>0$ and $%
\lambda_{11}<0$. In both cases this implies $\lambda_{12}=\sqrt{%
\lambda_{11}\lambda_{22}}$; (b) if $\lambda_{11}>0 \implies \lambda_{12}<0$
and $\lambda_{22}>0$ or if $\lambda_{22}>0 \implies \lambda_{21}<0$ and $%
\lambda_{11}>0$. In this case one has $\lambda_{12} =-\sqrt{%
\lambda_{11}\lambda_{22}}$. For both (a) and (b) cases we will have 
\begin{equation}
\bar\phi_1^2=-\frac{\lambda_{12}}{\lambda_{11}}\bar\phi_2^2 \implies
\bar\phi_1 = \pm {\biggl(\frac{\lambda_{22}} {\lambda_{11}}\biggr)}^{1/4}
\bar\phi_2  \label{eqreta}
\end{equation}
Also $W(\bar\phi_1,\bar\phi_2)=0$ and there is no kink-like solutions
connecting any points in the lines of minimum energy.

The next study concerns the coalesced ellipses of minima, which is depicted
in Fig~1.(c). In this case we have the trivial $(0,0)$ solution plus a
continuum of minima represented by the degenerated ellipse. We have $W(0,0)=0
$, $W(\pm \phi _{1}^{\ast },0)=(1/4)(\mu _{1}^{2}/\lambda _{11})$, $W(0,\pm
\phi _{2}^{\ast })=(1/4)(\mu _{2}^{2}/\lambda _{22})$ and $W(\pm \phi
_{1}^{\ast },0)=W(0,\pm \phi _{2}^{\ast })$. This means a null energy for
all orbits connecting the coalesced ellipses. The energy of a kink-like
structure connecting a point from the ellipse and the origin $(0,0)$ is
given by $E=|W(0,0)-W(\bar{\phi}_{1},\bar{\phi}_{2})|=(1/4)(|\mu
_{1}^{2}/\lambda _{11}|)$.

\begin{figure}[ht]
\begin{center}
\includegraphics[width=14cm]{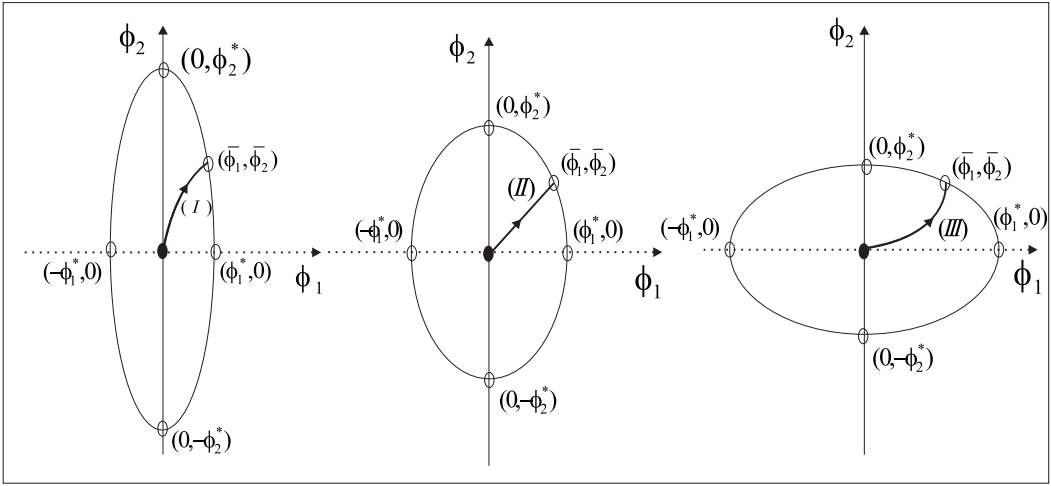} 
\end{center}
\caption{Phase space for $\protect\mu_1/\protect\mu_2=\protect\lambda_{11}/%
\protect\lambda_{12}$; $\protect\lambda_{12}^2=\protect\lambda_{11}\protect%
\lambda_{22}$ showing orbits connecting the point $(0,0)$ to the coalesced
ellipses. Orbit (I) is for $\protect\lambda_{11}/\protect\lambda_{22}>1$,
(II) for $\protect\lambda_{11}/\protect\lambda_{22}=1$ and (III) is for $0<%
\protect\lambda_{11}/\protect\lambda_{22}<1$.}
\label{coalescellip}
\end{figure}

The trial orbits method can be used to find an explicit solution  for $%
\phi_1(x)$ and $\phi_2(x)$ that connects  $(0,0)\rightarrow
(\bar\phi_1,\bar\phi_2)$. We try a solution of the form  
\begin{equation}
\phi_1 = A\phi_2^B  \label{trial}
\end{equation}
To satisfy the minimum energy points one must have  $A=\bar\phi_1/\bar\phi_2^B$%
.  Differentiating Eq. (\ref{trial}) we obtain 
\begin{equation}
\phi_1^{\prime}=AB\phi_2^{B-1}\phi_2^{\prime}\implies \frac{\phi_1^{\prime}}{%
\phi_1} = B \frac{\phi_2^{\prime}}{\phi_2}  \label{propellipses}
\end{equation}
But, considering equations \eqref{genmoveq} we see that this is equivalent
to a proportional relation among the two ellipses, in the non-degenerated
case. We can obtain the B parameter after substituting explicitly the
equations of the ellipses in Eq. (\ref{propellipses}). This gives 
\begin{equation}
\phi_1={\bar\phi_1}\biggl(\frac{\phi_2}{\bar\phi_2}\biggr) ^{\sqrt{%
\lambda_{11}/\lambda_{22}}}  \label{orbitc}
\end{equation}
and the structure of the orbit depends strongly on the product $%
\lambda_{11}\lambda_{22}$, as shown in Fig.~\ref{coalescellip}.

\begin{figure}[ht]
\begin{center}
\includegraphics[height=6cm]{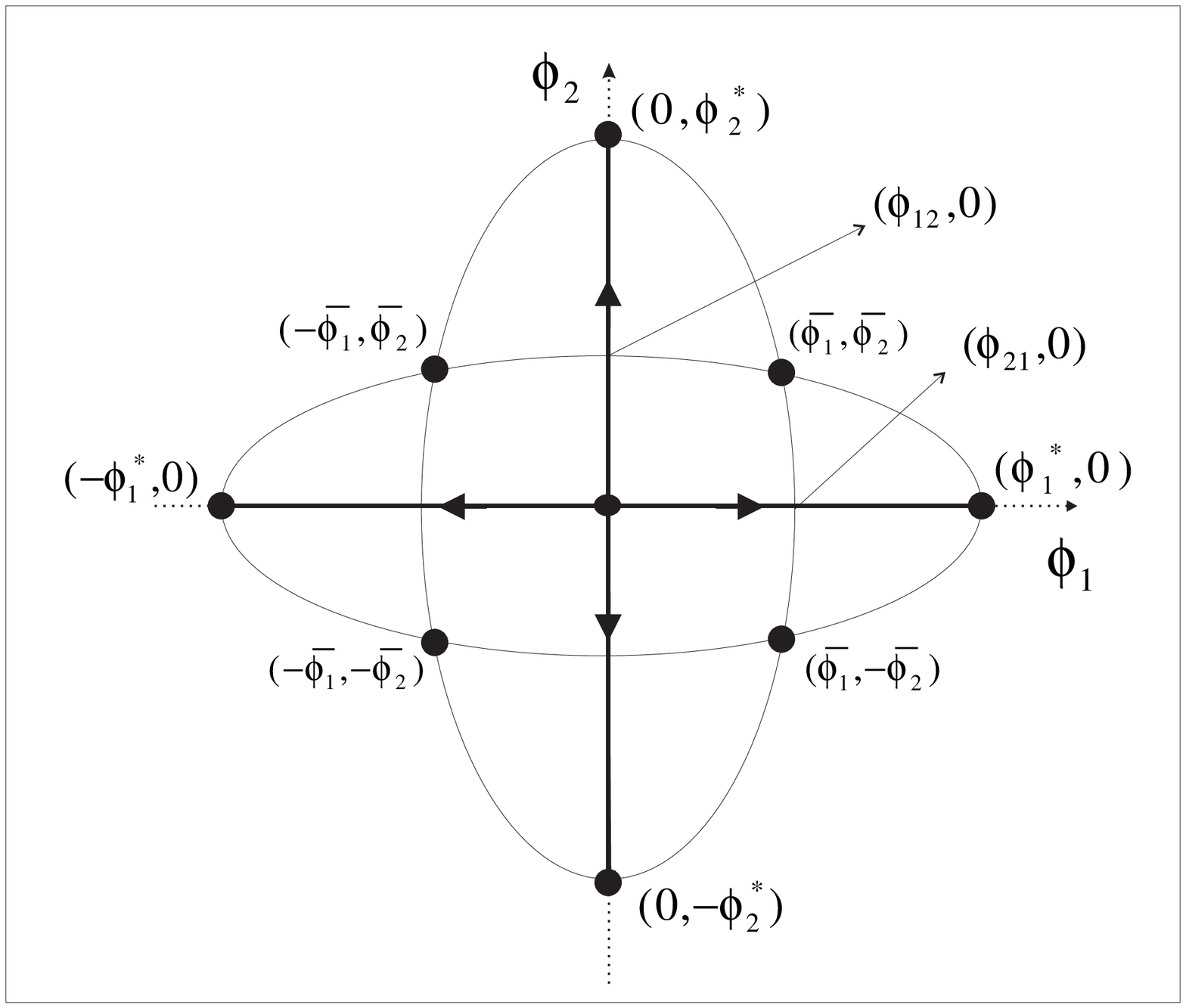} %
\includegraphics[height=6cm]{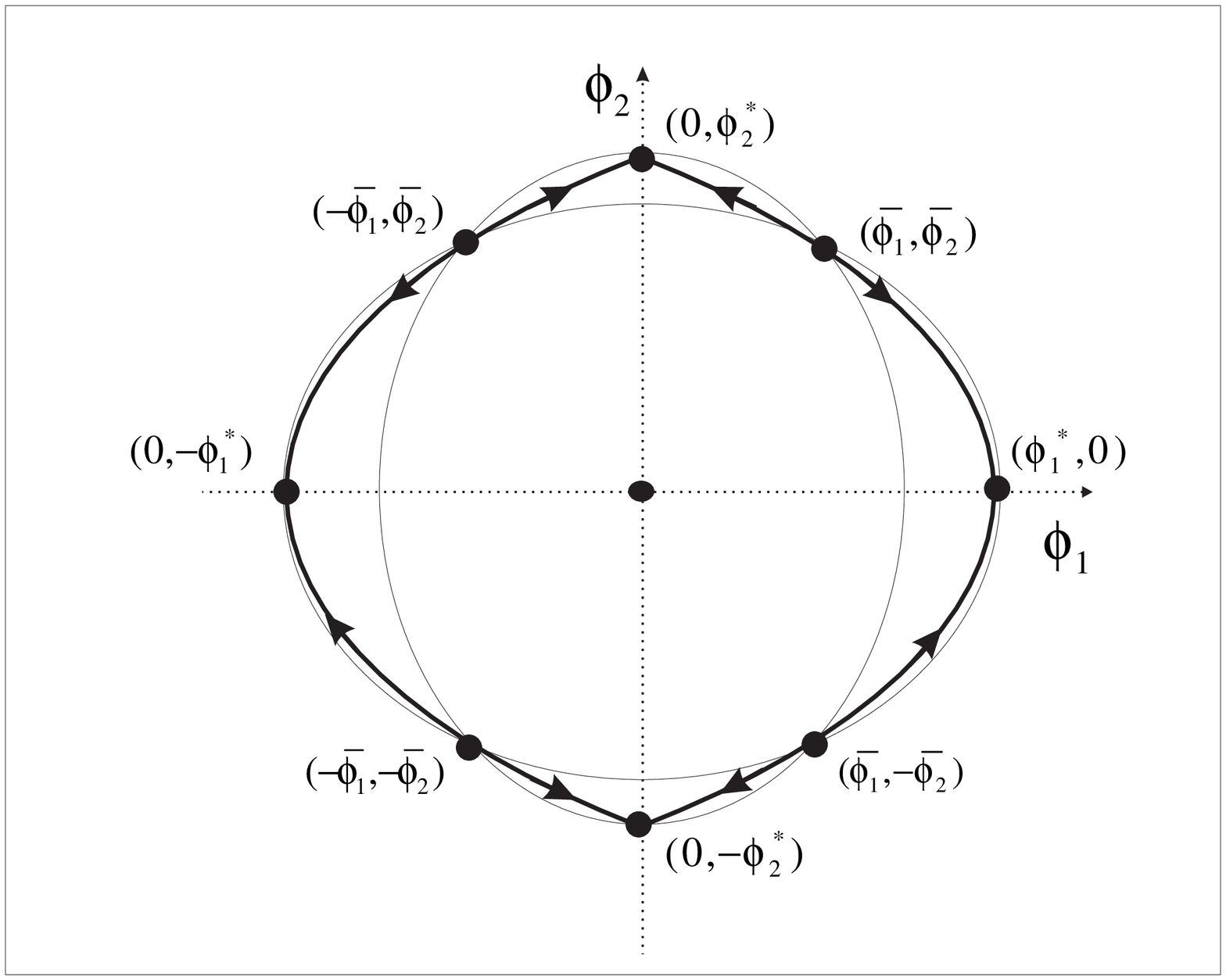}
\end{center}
\caption{(a) Phase space for $\protect\mu_1/\protect\lambda_{12}, \protect\mu%
_2/\protect\lambda_{21}>0;$ $\protect\mu_1/\protect\lambda_{11}, \protect\mu%
_2/\protect\lambda_{22}>0$ showing the 9 minimum energy points - 4 of them
obtained by the intersection points of the ellipses - and some orbits. (b)
Phase space for $\protect\lambda_{11}=2\protect\lambda_{12}$ and $\protect%
\lambda_{11}/\protect\lambda_{22}=4+\protect\mu_1/\protect\mu_2$, showing
the orbits $\protect\phi_1=\pm\protect\phi_1^*(-{\protect\phi_2^2}/{{\protect%
\phi_2^*}^2}+1) .$}
\label{intersecellip}
\end{figure}

We now deal with the case of intersecting ellipses, which is depicted in
Fig.~\ref{alldiagrams}(a). This case is very interesting, and it is better
to refer to Fig.~\ref{intersecellip}a, which shows the general configuration
for the minimum energy points, where we defined $\phi_{12}\equiv\sqrt{%
\mu_1/\lambda_{12}}$ and $\phi_{21}\equiv\sqrt{\mu_2/\lambda_{21}}$.

To obtain the energies of the solutions connecting minima we first consider
Eq.~(\ref{Wfichi}). We have, by symmetry, $W(\phi_1,\phi_2)=W(\phi_1,-%
\phi_2)=W(-\phi_1,\phi_2)$ $=W(-\phi_1,-\phi_2)$; thus, there are no
kink-like structure connecting the intersecting points from the ellipses.
Also, we have $W(\phi_1^*,0)$ $=W(-\phi_1^*,0)$ $= |\mu_1^2/(4\lambda_{11})|$
and $W(0,\phi_2^*)$ $=W(0,-\phi_2^*) =|\mu_2^2/(4\lambda_{22})|$. We studied
the following cases:

i) One connection by means of a straight line between $(0,0)\rightarrow
(0,\pm\phi_2^*)$, with energy $E_1=|W(0,0|-W(0,\pm\phi_2^*)|=|\mu_2^2/(4%
\lambda_{22})|$. This can be found solving the equations of motion to obtain 
\begin{equation}
\phi_1(x)=0, \,\,\,\, \phi_2(x)=\pm\phi_2^* \left(\frac{1+\tanh{(\mu_2 x)}}{2%
}\right)
\end{equation}
This solution represents a laser operating only on mode 2, where the laser intensity smoothly increases from zero to the maximum operating value.
By symmetry one can easily find similar solutions that connects $%
(0,0)\rightarrow (\pm\phi_1^*,0)$ where the laser operates only on mode 1.

ii) We look for solutions that connect $(0,0)\rightarrow
(\bar\phi_1,\bar\phi_2)$, with energy $E_3=|W(0,0|-W(\bar\phi_1,\bar\phi_2)|$%
. We try the orbit 
\begin{equation}
\frac{\phi_1}{\bar\phi_1}=\biggl(\frac{\phi_2}{\bar\phi_2}\biggr)^{B}
\end{equation}
Differentiating the orbit and substituting the first order equations leads
to 
\begin{equation}
\mu_{1} - \lambda_{11} \phi_1 ^{2} -\lambda_{12} \phi_2^{2} = B
(\mu_{2}-\lambda_{21} \phi_1^{2} - \lambda_{22} \phi_2 ^{2})
\end{equation}
Equating the independent coefficients we obtain $B={\mu_1}/{\mu_2}$ and the
remaining condition can be written as 
\begin{equation}
\biggl(\lambda_{11}-\frac{\mu_1}{\mu_2}\lambda_{21}\biggr)\phi_1^2= \biggl(%
\frac{\mu_1}{\mu_2}\lambda_{22}-\lambda_{12}\biggr)\phi_2^2
\end{equation}
\begin{equation}
\bar\phi_1^2 \det\Lambda^{(\phi_2)} \biggl(\frac{\phi_2^2}{\bar\phi_2^2}%
\biggr)^B= \bar\phi_2^2 \det\Lambda^{(\phi_1)} \frac{\phi_2^2}{\bar\phi_2^2}
\end{equation}
We then have the following possibilities:

$\bullet$ $\det\Lambda^{(\phi_1)}=\det\Lambda^{(\phi_2)}=0$. This leads to
the already analyzed case of coalesced ellipses.

$\bullet$ $\det\Lambda^{(\phi_1)},\,\det\Lambda^{(\phi_2)}\neq 0$. This
leads to $B=1$ and $\bar\phi_1^2 \det\Lambda^{(\phi_2)} = \bar\phi_2^2
\det\Lambda^{(\phi_1)}$, which means $\mu_1=\mu_2$ and $\lambda_{11}=%
\lambda_{22}\equiv\lambda$, respectively, with $\phi_1=\phi_2$. In this case
one can obtain an expression for $\phi_1(x)$ by means of substituting the
former equation into the equation of motion for the field $\phi_1$(cf. Eq.[%
\ref{genmoveq1}]). One finds 
\begin{equation}
\phi_1(x)=\phi_2(x)=\pm\sqrt{\frac{\mu} {\lambda_{12}+\lambda}\cdot\frac{1}{2%
} (1+\tanh(\mu x))}
\end{equation}
One can see that for $x\rightarrow -\infty$, $(\phi_1,\phi_2)\to(0,0) $ and
for $x\rightarrow +\infty$, $(\phi_1,\phi_2)\to(\pm\bar\phi_1,\pm\bar\phi_2)$%
. This corresponds to a laser operating is both modes with the same intensity, since we have for this solution $\phi_1(x)^2=\phi_2(x)^2$, which corresponds to $E_1(t)^2=E_2(t)^2=I(t)$. The intensity of the i-th mode increases continuously  until achieving the maximum value given by $I_{max}={\mu}/{(\lambda_{12}+\lambda)}$.

iii) For the orbit $\phi_1=A\phi_2^2+B$ to connect $(\phi_1^*,0)\to(0,%
\phi_2^*)$, one must have $B=\sqrt{\mu_1/\lambda_{11}}=\phi_1^*$ and $%
A=-(\lambda_{22}/\mu_2)\sqrt{(\mu_1/\lambda_{11})}$ $= - \phi_1^*/{\phi_2^*}%
^2$. Then we have the orbit 
\begin{equation}
\frac{\phi_1}{\phi_1^*}+\biggl(\frac{\phi_2}{\phi_2^*}\biggr)^2=1
\end{equation}
Deriving the orbit and using equations \eqref{genmoveq} leads to the
consistency conditions 
\begin{equation}
\lambda_{11}=2\lambda_{12} , \,\,\,\, \frac{\lambda_{11}}{\lambda_{22}}=4+2%
\frac{\mu_1}{\mu_2}  \label{nu2lambda}
\end{equation}

We can show that conditions (\ref{nu2lambda}) are also compatible with the
Eqs. (\ref{barphi}) and (\ref{barchi}) that define the intersecting points.
Also, the condition for existence of the minimum energy points $%
(\bar\phi_1,\bar\phi_2)$ leads to another constraint on the parameters. We
can see in Fig.~\ref{intersecellip}(a) that crossing among the ellipses
exist only if $\phi_1^*\geq \phi_{21}$ and $\phi_2^*\geq\phi_{12}$. This
leads after using the consistency conditions to $2\leq\mu_1/\mu_2\leq
2+\mu_1/\mu_2$. This inequality is satisfied only when $\mu_1/\mu_2\geq 2$.
When the ellipses do not cross one another, we must have, as one
possibility, that $\phi_1^*\geq \phi_{21}$ and $\phi_2^*\leq\phi_{12}$.
This, with the conditions (\ref{nu2lambda}) lead to the condition $%
\mu_1/\mu_2\geq 2+\mu_1/\mu_2$ which is an impossibility. So, this type of
orbit needs the crossing among the ellipses. The phase space diagram for
this orbit is shown on Fig.~\ref{intersecellip}(b). There one shows that the
points $(\pm\bar\phi_1,\pm\phi_2)$ are unstable. In this way the orbits connect one of these points, for $x\to -\infty$, to 
one of the other minimum energy points $(0,\pm\phi_1^*)$ or $(\pm\phi_2^*,0)$, for $x\to\infty$.

As an example we can choose $\lambda_{11}=3,\lambda_{22}=1/4,\mu_1=3,%
\mu_2=3/4$. This means a ratio $\mu_1/\mu_2=4\geq2$ and $\lambda_{12}=%
\lambda_{11}/2=3/2$. This choice corresponds to an initial condition where mode 1 (represented by the $\phi_1$ field) is well above threshold, whereas mode 2 (corresponding to the $\phi_2$ field) has a smaller gain.
We have minimum energy points $(\phi_1^*,0)=(1,0)$, $%
(\bar\phi_1,\pm\bar\phi_2)=(1/2,\pm\sqrt{6}/2)$, $(0,\pm\phi_2^*)=(0,\pm%
\sqrt{3})$ and an orbit $\phi_1=-(1/3)\phi_2^2+1$ connecting theses 6 points
(three for $\phi_2\geq0$ and three for $\phi_2\leq0$). Substituting the
orbit in Eq.~\eqref{genmoveq1}, we obtain 
\begin{equation}
\frac1{\phi_1}\frac{d\phi_1}{dx}=-\frac{3}{2}(2\phi_1^2-3\phi_1+1)
\label{sqrt partic}
\end{equation}
Integrating the former equation we obtain two solutions that agree with $%
\phi_2\geq0$, namely  ($(\phi_1^+(x),\phi_2^-(x))$ and $(\phi_1^-(x),%
\phi_2^+(x))$, with: 
\begin{equation}
\phi_1^{\pm}(x)=\frac{1}{2} \left(1\pm\sqrt{\frac{1}{2}(1+\tanh(3x/4)}\right)
\end{equation}
and 
\begin{equation}
\phi_2^{\mp}(x)=\sqrt{\frac{3}{2} \left(1\mp\sqrt{\frac{1}{2}(1+\tanh(3x/4)}%
\right)}
\end{equation}
For $(\phi^{+}_1(x),\phi^{-}_2(x))$ we have $\lim_{x\to-\infty}\phi^+_1=1/2$%
, $\lim_{x\to-\infty}\phi^-_2=\sqrt 6/2$ and $\lim_{x\to+\infty}\phi^+_1=1$, 
$\lim_{x\to+\infty}\phi^-_2=0$ and the orbit connects $(\bar\phi_1,\bar%
\phi_2)\rightarrow(\phi_1^*,0)$ as $(1/2,\sqrt6/2)\rightarrow(1,0)$. Here we have a final state where only mode 1 oscillates. In this regime we can say that oscillation in mode 2 was quenched \cite{nuss} by the oscillation in mode 1. For $%
(\phi^{-}_1(x),\phi^{+}_2(x))$ we have $\lim_{x\to-\infty}\phi^-_1=1/2,%
\lim_{x\to-\infty}\phi^+_2=\sqrt 6/2$ and $\lim_{x\to+\infty}\phi^-_1=0,%
\lim_{x\to+\infty}\phi^+_2=\sqrt3$, and the orbit connects $%
(\bar\phi_1,\bar\phi_2)\rightarrow(0,\phi_2^*)$ as $(1/2,\sqrt6/2)%
\rightarrow(0,\sqrt3).$ Now we have the opposite regime, where mode 2 absorbs energy continuously and mode 1 decreases until only mode 2 remains.

In conclusion, in this work we have used the trial orbit method introduced
by Rajaraman \cite{t4} to investigate first order differential equations
which appear when one uses the Bogomol'nyi approach to study minima energy
kink-like solutions \cite{bog}. We have studied a family of models described
by two real scalar fields inspired on the theory of two-mode laser. We have
determined all the minimum energy points in terms of the parameters which
specify the model. We have found a rich structure of minima, and several
analytical solutions of the kink-like type, connecting pairs of minima in
the field space. 
In order to correctly map our results for $\phi_1(x)$ and $\phi_2(x)$ to  
the time-dependent problem of the dynamical competition between the two modes $E_1(t)$ and $E_2(t)$, we must interpret $x$ from our mathematical solutions as the physical time $t$. This is justifiable after comparing Eqs. (\ref{genmoveq1})-(\ref{genmoveq2}) from our classical field theory with Eq. (\ref{eqEn}) from the semiclassical laser theory. In this way some of the exact solutions here studied where used to study phenomenological situations such as laser oscillations between two
modes in a multi-mode system.

The authors would like to thank CAPES, CNPq, MCT-CNPq-FAPESQ and
MCT-CNPq-CT-Infra for partial support. The authors also thank C. Furtado for discussions and M. M. Ferreira
Jr. for reading the manuscript.



\end{document}